\documentclass[11pt,a4paper,notoc,twoside]{JHEP3}

\usepackage{graphicx,epsfig,rotating}
\usepackage{amssymb}
\usepackage{latexsym}
\usepackage{citesort}
\DeclareSymbolFontAlphabet{\mathbb}{AMSb}

\newcommand{\veckext}{\mbox{${\vec k}^{(ext)}$}}
\newcommand{\veck}{\mbox{${\vec k}$}}
\newcommand{ \CYB}{\mbox{$CY{\mathbb B}$}}

\title{ Root systems from Toric Calabi-Yau Geometry.
Towards new algebraic structures and symmetries in physics?}
\preprint{\small  CERN-PH-TH/2004-132\\
IFT-UAM/CSIC-04-06\\
UM-FT/04-67}
\author{ E. Torrente-Lujan${}^{1\star}$,  G. G. Volkov${}^{2\star}$\\
{\small
$^1${\it GFT, Dept. of Physics, Universidad de Murcia,Spain}\\
$^2${\it IFT, Univ. Autonoma de Madrid, Cantoblanco, Madrid, Spain, \\ on leave from PNPI, Gatchina, St 
Petersburg,Russia}
}\\
\email{ e.torrente@cern.ch},\email{ guennadi.volkov@cern.ch}}
\abstract{
 The algebraic approach to the construction of the reflexive polyhedra that
yield Calabi-Yau spaces in three or more complex dimensions with K3 fibres
reveals graphs that include and generalize
 the Dynkin diagrams associated
with gauge symmetries. 
In this work
we continue to study the structure of graphs obtained
 from $CY_3$
reflexive polyhedra. We show how some particularly 
defined integral matrices can be assigned to these diagrams. 
This family of matrices and its associated  
 graphs may be obtained by
relaxing the restrictions on the individual entries 
 of the
generalized Cartan matrices associated with the Dynkin diagrams that
characterize Cartan-Lie and affine Kac-Moody algebras. 
These graphs keep however the affine structure, as it was in  Kac-Moody Dynkin
diagrams. 
We presented a possible root structure for some 
simple cases. 
We conjecture that
these generalized graphs and associated 
link matrices may characterize generalizations of these
algebras. 
}


\begin{document}

\section{Introduction}

The final objective of this work is no short of ambition: to discuss the question of what 
is nature of the ultimate, or at least of the next, set of symmetries 
of the Standard Model.
The success of the use of  Cartan-Lie Algebras and their direct 
products in the GSW model got full confirmation by the discovery 
of the intermediating bosons via the observation of neutral currents and 
is now beyond any doubt.
However the efforts dedicated to the extension of the SM group of 
symmetries, always in the frame of Cartan-Lie algebras and with the 
objective of, for example, reducing the number of free parameters appearing 
in the theory, has not lead to the same level of success.
Equally, the attempts of Grand Unified theories to englobe 
the direct product of the symmetries 
of the SM in some larger group has become problematic, with not
clear new predictions and with a systematic  too large
proton instability.

At a very basic level, and without any obvious direct interest to SM, 
Cartan-Lie symmetries are closely connected to 
the 
geometry with symmetric homogeneous spaces whose classification was 
performed by Cartan himself in the classic epoch. 
Furtherly, the developing of an alternative 
geometry of non-symmetric spaces appeared later, their classification 
was suggested in 1955 by Berger using holonomy theory \cite{Berger}:  
there are  some infinite series of spaces with holonomy groups 
$SO(n)$, $U(n)$, $SU(n)$, $Sp(n)\times Sp(1)$, $Sp(n)$ 
and, in addition, some exceptional
spaces  with holonomy groups $G(2)$, $Spin(7)$, $Spin(16)$.

The Standard Model does not provide, at least visibly to us, any clue 
on how to attack the problem of the nature of symmetries 
at this  very basic geometry level. 
This is not the case for the superstring theories and general Kaluza-Klein
scenarios, for example 
the compactification
of the heterotic string leads to the classification of states in a 
representation of the Kac-Moody algebra of the gauge group
$E_8\times E_8$ or $Spin(32)/{ Z}_2$. In another words,
 heterotic superstring discovered for physics
the 6-dimensional Calabi-Yau space, an example of  non-symmetrical space 
having the  $SU(3)$ group of holonomy \cite{CHSW}.
It has also been shown \cite{belavin} that group and algebra theory appear at the 
root of generic two-dimensional conformal field theories (CFT). 
The basic ingredient 
here is the central extension of infinite dimensional Kac-Moody algebras.
There is a clear  connection between these algebraic and geometric 
generalizations. Affine Kac-Moody algebras are realized as the central 
extensions of Loop algebras: the set of mappings on a compact manifold, 
for example $S^1$, which take value on a finite-dimensional Lie algebra.
In summary, Superstring theory intrinsically contains a number of 
other infinite-dimensional algebraic symmetries such as the Virasoro 
algebra associated with conformal invariance and to generalizations 
of Kac-Moody algebras themselves, such as hyperbolic and Borcherd algebras.

%


Remaining with Calabi-Yau spaces, Dynkin diagrams 
(or Coxeter-Dynkin diagrams) which are in one-to-one correspondence 
not only to Cartan-Lie but also Kac-Moody algebras 
has also been  observed through the technique of resolution 
of singularities.
The rich singularity structure of some examples of 
non-symmetrical Calabi-Yau 
spaces gives us here 
another opportunity to re-obtain infinite dimensional, affine,
 Kac-Moody symmetries. 
The Cartan matrix of affine Kac-Moody groups is identified with the 
intersection matrix of the union of the complex proyective lines resulting 
from the blow-up of the singularities.
This is indeed the case of $K3 \equiv CY_2$ where the  classification of 
 degeneration of their fibers and their associated singularities, leads us 
to  link 
  $CY_2$ spaces with the infinite and exceptional 
 series  $A_r^{(1)}$, $D_{r}^{(1)}$,  
$E_6^{(1)}$, $E_7^{(1)}$, $E_8^{(1)}$ of affine Kac-Moody algebras.

The study of Calabi-Yau spaces appearing in superstring, F and M theories 
can be approached from the theory of Toric geometry and the so 
called Batyrev's construction  \cite{Bat} where the concept
of reflexive polyhedra appears.
This concept of reflexivity or mirror symmetry has been 
linked  \cite{CF}  to the problem of the duality  between 
superstring theories compactified in different (K3 and CY3) Calabi-Yau 
spaces.
The same Batyrev construction has also been used to show  
how subsets of points in these reflexive polyhedra can be 
identified with the Dynkin 
diagrams ~\cite{CF,CPR,Greene,KV} 
of the affine version of the 
gauge groups appearing in superstring and F-Theory. More explicitly, 
the gauge content of the compactified 
theory can be read off from the  {\em dual}
reflexive polyhedron  of the Calabi-Yau space which is used 
for the compactification.

For a K3 Calabi-Yau space, subdivision of the reflexive polyhedron 
in different subsets separated by polygons which are themselves 
reflexive is equivalent to establish a fibration structure for the 
space. The fiber simply being the corresponding one to the 
this intersecting mirror polygon.
Any of these parts of the full polyhedron has been 
called by Candelas and Font  a ``top''\cite{CF}. 
Thus, reflexive polyhedra  when intersected 
with a plane gives raise to individual reflexive polygons.
This intersection allows us to define two parts of the polyhedron: 
``top'' and ``bottom'' in the nomenclature of Ref.\cite{CF}
 or left and right in what follows in this work.

 Calabi-Yau spaces being
characterized geometrically by reflexive { Newton } polyhedra can be 
enumerated systematically. However, one might going 
deeper than simple enumeration,
 it has been recently realized that
different 
reflexive polyhedra are related algebraically via what it has been termed
Universal Calabi-Yau Algebra (UCYA). The term ``Universal'' comes 
from the fact that  includes, beyond binary, ternary and
higher-order operations.

The UCYA is particularly well suited for exploring fibrations of 
Calabi-Yau spaces, which are visible as lower-dimensional slices through 
higher-dimensional mirror polyhedra or, alternatively, as 
projections on the reflexive polyhedron. The UCYA approach naturally 
gives these slice or projection structures.
See Fig.~1 in Ref.\cite{Vol} and its explanation for 
a description of the elliptic fibration of a K3 space.
The Left and Right parts of the reflexive polyhedron corresponds
correspond each to a, so called, extended vector.
According to the  UCYA scheme, 
the sum, a binary operation,
 of these  two ``extended vectors'' gives true reflexive vector,
describing the full $CY_2 =K3$ manifold. 
This means that 
a direct algebraic relation between  K3( = CY$_2$)
and CY$_3$ spaces is established. This property is completely general:
it has  been  shown previously how the UCYA, with its rich 
structure of beyond binary operations,  can be
used to generate and interrelate  $CY_n$ spaces of any order.
 The UCYA provides in addition 
a complete and systematic description of analogous
decompositions or nesting of fibrations in  Calabi-Yau spaces of 
any dimension.

One of the remarkable features of Fig.~1 in Ref.\cite{Vol} is that
each of the right and left set of nodes 
 constitute graphs corresponding to  
affine Dynkin diagrams: the  $E_6^{(1)}$ and  $E^{(1)}_8$ diagrams. 
This is not a mere coincidence or 
an isolated example. 
As discussed there, 
 all the elliptic fibrations of K3 spaces 
found using the UCYA construction 
feature this decomposition into a pair of graphs that 
can be interpreted as Dynkin diagrams.


The purpose of this paper is to deal with the 
 important problem on how to generalize previous results,
applicable to K3 only, to any Calabi-Yau space with any dimension 
and any fiber structure.
In first place, we would like  to know 
which  Dynkin-like  diagrams can be found ``digging'' in higher 
dimension Calabi-Yau spaces. As it  has firstly been 
shown in Ref.\cite{Vol} a bounty of new diagrams,
``Berger Graphs'', can be brought to light by this procedure.
Our hypothesis is that these graphs correspond, in some undefined 
way, to some new algebraic structure, as Dynkin diagrams are in 
one-to-one correspondence with root systems and Cartan matrices in 
 semi-simple Lie Algebras and affine Kac-Moody algebras.

Based on this,  our next objective is to  
 construct a similar theory to Kac-Moody algebras where some newly 
defined generalized Cartan 
matrix fulfilling extended conditions are introduced. 
There are plenty of possible, trivial 
and not so trivial, generalizations of Cartan matrices where the rules for 
the diagonal and non diagonal entries are modified, it is impossible 
to find all of them and classify. On the other hand, probably not all them 
give meaningful, consistent  generalizations of Kac-Moody algebras and 
probably even less of them have interesting implications for physics. 
One has to find natural, and hopefully physically inspired,
 conditions on these matrices. The relation of Berger Graphs to Calabi-Yau 
spaces could be this physical inspiring link.

Having the equivalent of Cartan Matrix we can use standard algebraic 
tools, definition of a inner product, construction of root systems and their group of transformation etc, which could be helpful in clarifying the 
meaning and significance 
of this construction.

The structure of this paper is as follows.
In section 2 we show how to extract graphs directly from the 
polyhedra associated to Calabi-Yau spaces and how, from these 
graphs define new ones by adding or removing nodes.
We include a first step towards the systematic enumeration of 
rules to build generalized Cartan matrices from the graphs,
from this we give an algebraic definition of a new set of matrices 
generalizing the Cartan-Kac Moody matrices and, finally, rules 
for the   reverse process of obtaining graphs from matrices.
In section 3 we present some elementary but illustrative 
examples of these procedures.
Finally we draw some conclusions and conjectures.

\section{UCYA and generalized Dynkin diagrams.}

One of the main results in the Universal Calabi-Yau Algebra (UCYA)  is
that the reflexive weight vectors (RWVs) $\vec{k_n}$ of dimension $n$,
which are the fundament for the construction of CY spaces,
 can
obtained directly from lower-dimensional RWVs $\vec{k_{1}}, \ldots,
\vec{k_{n-r+1}}$ by algebraic constructions of arity
$r$~\cite{AENV1,AENV2,AENV3,AENV4}. The dimension of the corresponding 
vector is $d+2$ for a Calabi-Yau $CY_d$ space.

For example, the sum of vectors,  
a binary composition rule of the UCYA, gives complete information about
the $(d-1)$-dimensional slice structure of $CY_d$ spaces.
In the K3 case, the 
Weierstrass fibered 91 reflexive weight vectors of the total of 95  $\vec{k_4}$ can be 
obtained by such binary, or arity-2, constructions out of just 
five RWVs of dimensions 1,2 and 3.

In a iterative process, we can combine by the same 2-ary operation the five vectors of dimension $1,2,3$ 
with these other 95 vectors 
to obtain a set of 4242 chains of 
 five-dimensional RWVs $\vec{k_5}$ 
CY$_3$ chains.
This process is  summarized
in Fig.~3 in Ref.\cite{Vol}.
By construction, the corresponding mirror 
CY$_3$ spaces are seen to possess K3 fibre
bundles.
 In this case, reflexive 4-dimensional polyhedra
are also separated into three parts: a reflexive 3-dimensional
intersection polyhedron and `left' and `right' skeleton graphs. 
The complete  description of a Calabi-Yau space
  with all its non-trivial $d_i$ fibre structures
needs a full range of n-ary operations where $n_{max}=d+2$.

It has been shown
in the toric-geometry approach
how the Dynkin diagrams of affine Cartan-Lie algebras appear in reflexive
K3 polyhedra
~\cite{CF,CPR,Greene,KV,Bat}. 
Moreover, along the same lines
using
examples of the lattice structure of reflexive polyhedra for CY$_n:  n
\geq 2$ with elliptic fibres, it has also been shown~\cite{AENV1,AENV3,AENV4}, 
 that there is a correspondence
 between the five basic RWVs
(basic constituents of composite RWVs describing  K3 spaces, 
see section 2 in \cite{Vol}) 
and affine Dynkin diagrams for the five ADE types of Lie
algebras ( A, D  series and exceptional E$_{6,7,8}$).

In each case, a pair of extended RWVs have an
intersection which is a reflexive plane polyhedron; each vector from
the  pair gives the left or right part of the three-dimensional RVW.
  The construction generalizes to any dimension.
In Ref.\cite{Vol}  it was remarked that in the corresponding 
``left'' and right ``graphs'' of 
$CY_{3,4,..}$ Newton reflexive polyhedra one can find 
new graphs with some regularity in its structure.


In principle one should be able to 
 build , classify and understand these regularities 
of the graphs according to the n-arity operation which originated the 
construction.
For the case of binary or arity-2 constructions: two graphs are possible.  
In general for any reflexive polyhedron,  
for a given   arity-r  intersection, it corresponds exactly $r$ graphs.

In the binary case, 
the 2-ary intersection (a plane) in the Newton polyhedra, which 
correspond to the {\em eldest} reflexive vector of the series,
 separate 
left and right graphs.
A concrete rule for the extraction of individual  graph points
from all possible nodes in the graphs 
is that they are selected if they exactly 
 belong ``on the edges''  lying on one side or another with respect the
intersection. 
In the ternary case, 
the  3-ary intersection hypersurface is a volume, which 
 separate three domains in the 
newton polyhedra and three graphs are possible. 
Individual points are assigned to each graph 
looking at their position with respect to 
 the volume intersection (see Tab.1 in \cite{Vol} for 
some aclaratory examples).

These are graphs directly obtained from the reflexive polyhedron construction.
On a later stage, we will define graphs independently of this construction.
These graphs will be derived, or by direct manipulation of them, or    from 
generalized Cartan matrices in a purely algebraic fashion. They will basically
consist on the primitive graphs extracted from reflexive polyhedra to whom 
internal nodes in the edges will have been added or eliminated.
The nature of the relation, if any, of the graphs thus generated 
to the  geometry of Toric varieties and 
the description of Calabi-Yau as hypersurfaces on them 
is  related to the possibility 
of defining viable ``fan'' lattices. This is an open question, 
clearly 
related  to the properties of the generalized Cartan matrices, interpreted 
as a matrix of divisor intersections. 
The consideration of these matrices 
is the subject of the next section.

\subsection{From Berger graphs and Dynkin diagrams to Berger matrices}

In the previous section we have
 established the existence of Dynkin-like graphs, possibly 
not corresponding to any of the known Lie or affine Kac-Moody algebras.
The information contained in the graphs can be encoded in a more 
workable structure: a matrix of integer numbers to be defined.
If  these  ``Dynkin'' graphs are somehow related 
to possible generalizations of the
 Lie and affine Kac-Moody algebra concepts, 
it is then natural to look for 
possible generalizations of the corresponding affine Kac-Moody Cartan matrices
when searching for possible ways of assigning integral matrices to them.

One possibility which could serve us of guide 
is to suppose that this affine property remains:
matrices with determinant equal to zero and all principal minors positive.  
We will see in what follows that this is a sensitive choice, on the other 
hand it turn out that the 
usual conditions on the value of the diagonal elements has to be abandoned.

({\em Building Matrices from graphs.})
We assign to any generalized Dynkin diagram,
a set of  vertices and lines connecting them,
 a matrix, $B$, whose non diagonal elements are 
either zero or are negative integers. 
There are different possibilities, for non diagonal elements,
 considering for the moment
 the most simple case of ``laced'' graphs leading to 
symmetric assignments, we have: 

\begin{itemize}

\item there is no line from the vertex $i$ to the vertex $j$. In this 
case the element of the matrix $B_{ij}=0$

\item there is a single line connecting $i-j$ vertices. In this case 
$B_{ij}=-1$
\end{itemize}
The diagonal entries should be defined in addition. 
As a first step, no special restriction is applied and any positive 
integer is allowed. We see however that very quickly  only a few 
possibilities are naturally selected. The diagonal elements of the matrix 
 are two for CY2 originated graphs but are allowed to take increasing 
integer numbers with the dimensionality of the space, 
$3,4...$  for $ CY_{3,4...}$.

We have checked (see also Ref.\cite{Vol}) a large number of 
graphs and 
matrices associated to them, obtained by inspection considering 
different possibilities.
Some regularities are quickly disclosed.
In first place it is easy to see that there are graphs where
the number of lines outgoing a determined vertex can be bigger than two,
 in  cases of interest
 they will be 3, graphs from CY3, or bigger in the cases of 
graphs coming from CY4 and higher 
dimensional spaces.
Some other important regularities appear. The matrices are genuine generalizations 
of affine matrices. 
Their determinant can be made  equal to zero and all their principal 
minors made positive by careful choice of the diagonal entries depending 
on the Calabi-Yau dimension and n-ary structure. 

Moreover, we can go back to the defining reflexive polyhedra and 
define  other quantities  in purely geometrical terms. For 
example  we can 
consider the position or distance of each of the  vertices of the 
generalized Dynkin diagram to the intersecting reflexive polyhedra.
Indeed, it has been remarked \cite{CF} that Coxeter labels 
for affine Kac-Moody algebras can be obtained 
directly from the graphs: they correspond precisely to this ``distance'' between 
individual nodes and some defined intersection which separates ``left'' and 
``right'' graphs.
Intriguingly, this procedure can be easily generalized to our case, one 
can  see that, by a careful choice of the entry assignment for the 
corresponding  matrix, it follows
 Coxeter labels can be given in a proper way: 
they have the expected 
property of corresponding to the elements of the null vector a generalized 
Cartan matrix.

From the emerging pattern of these  regularities, we are lead to
 define a new set of matrices, generalization 
of Cartan matrices in purely algebraic terms,
the Berger, or Berger-Cartan-Coxeter matrices.
 This will be done in the next paragraph.

({\em Definition of affine Berger Matrices}).
Based on previous considerations, we define, now in purely algebraic 
terms, the so called Berger Matrices.
We suggest the following rules for them, 
in what follows we will see step by step how they lead
to a consistent construction generalizing the  Affine Kac-Moody concept.
A Berger matrix is a finite integral matrix characterized by the following data:
\begin{eqnarray}
{\mathbb B}_{ii}&=&2,3, 4..\nonumber\\
{\mathbb B}_{ij}& \leq& 0,\quad 
{\mathbb B}_{ij} \in {\mathbb Z} ,\nonumber\\
{\mathbb B}_{ij}=0 &\mapsto & {\mathbb B}_{ji}=0, \nonumber\\
Det\ {\mathbb B} &=&0,\nonumber\\
Det\ {\mathbb B}_{\{(i)\}} &>& 0.\nonumber
\label{eqsberger}
\end{eqnarray}
The last two restrictions, the zero determinant and the positivity of all 
principal proper minors,  corresponds to the  {\it affine condition}. 
They are shared by Kac-Moody Cartan matrices, so we expect that the 
basic definitions and properties of those can be easily generalized.
However, with respect to them, 
we relaxed the restriction on the diagonal elements. Note that,
more than one type of diagonal entry is 
allowed: $2,3,..$ diagonal entries can coexist in a given matrix.
This apparently 
minor modification has in turn important consequences \cite{newtorrente}
 when we define a Weyl group for the theory.

For the sake of convenience, we define also 
{\em ``non-affine'' Berger Matrices}
where the condition of non-zero determinant is again imposed. These matrices 
does not seem to appear naturally resulting from polyhedron graphs 
but they are 
useful when defining root systems and Weyl group for the affine case by extension of them. They 
could play  the same role of basic simple blocks as finite Lie algebras play for the
case of affine Kac-Moody algebras.

The important fact to be remarked here is that this definition 
lead us to a construction with the right properties we would 
expect from a generalization of the Cartan matrix idea.

({\em  From any Berger matrix to a Berger-Dynkin diagram.})
The systematic enumeration of the various possibilities concerning the 
large family of possible Berger matrices can be facilitated 
by the introduction for each matrix of its generalized Dynkin diagram.
As we intend that the definition of this family of matrices be independent 
of algebraic geometry concepts we need an independent definition of these 
diagrams.  
Obviously the procedure given before 
can be reversed to allow the deduction of the 
generalized Dynkin diagram from its generalized Cartan or Berger Matrix.
An schematic  prescription for the most simple cases could be:
A) For a matrix of dimension $n$, define  $n$ vertices and  draw them
 as small circles. In case of appearance of vertices with different 
diagonal entries, some graphical distinction will be performed.
Consider all the element $i,j$ of the matrix in turn.
B) Draw one line from vertex $i$ to vertex $j$ if the corresponding 
element $A_{ij}$ is non zero.

In what follows, we show that indeed these kind of matrices and 
Dynkin diagrams, exist beyond those  purely defined from 
 Calabi-Yau newton reflexive polyhedra.
In fact we show that there are infinity families of them where 
suggestive regularities appear.

It seems easy to conjecture that 
the set of all, known or generalized,
 Dynkin diagrams obtained from Calabi-Yau spaces 
can be described by this set of Berger matrices. 
It is however not so clear the validity of the opposite   
question,
whether or not the infinite set of generalized Dynkin diagrams 
previously defined can be found digging in the Calabi-Yau $(n,a)$ structure 
indicated by UCYA.
For physical applications however it could be important the following 
remark. Theory of Kac-Moody algebras show us that for any 
{\em finite} or {\em affine} Kac-Moody
 algebra, every proper subdiagram 
(defined as that part of the 
generalized Coxeter-Dynkin diagram obtained by removing one or more vertices 
and the lines attached to these vertices) 
is a collection of diagrams corresponding to {\em finite} Kac-Moody algebras.
In our case we have more flexibility. Proper subdiagrams, obtained 
eliminating internal nodes or vertices, are in general collections of 
Berger-Coxeter-Dynkin diagrams corresponding to other (affine by construction )Berger diagrams 
{\em or} to {\em affine} Kac-Moody algebras.
This property might  open the way to the consideration of non-trivial 
extensions of SM and string symmetries.

({\em The Berger Matrix as an inner product: 
construction of root spaces.})
For further progress, the interpretation of a Berger matrix as the matrix of 
divisor intersections $B_{ij}\sim D_i\cdot D_j$ in Toric geometry
could be useful for the study of the 
viability of fans of points associated to them, singularity blow-up, and the existence of 
Calabi-Yau varieties itself. This geometrical approach will be pursued 
somewhere else \cite{newtorrente}.
However, for algebraic applications, and with the extension of the CLA and KMA 
concepts in mind, the interpretation of these matrices as matrices 
corresponding to a inner product in some vector space is most natural which 
is our objective now.

The Berger matrices are obtained by weaking the conditions on 
the generalized Cartan matrix ${\hat {\mathbb A}}$ appearing in affine 
Kac-Moody algebras.
In what concern algebraic properties, there are no changes, 
it remains intact the condition of semi-definite positiviness, 
this allows to translate trivially many of 
the basic ideas and terminology for roots and root subspaces 
for  appearing in Kac-Moody 
algebras. Clearly,
the problem of expressing the ``simple'' roots in a orthonormal 
basis was an important step in the classification of semisimple
Cartan-Lie algebras.

The apparent minor modification which has been introduced: 
to allow for diagonal entries different 
from two, has however deep impact in another place: when we define 
the generalization of the Weyl group of transformations, we easily realize 
that this is in general an infinite group enlarged, with  respect 
the also infinite Weyl group appearing in affine Kac-Moody case, by 
a new infinite set of transformations. 
The detailed study of the properties of these set of transformations will 
be the object of a next study elsewhere \cite{newtorrente}.

For a Berger matrix $ \mathbb B_{ij}$ of dimension $n$, the rank is $r=n-1$.
The  $ (r+1)\times (r+1) $ dimensional is nothing else that a 
generalized Cartan matrix. 
We can suppose without loss of generality this matrix to be symmetric 
(if we start with a non-symmetric Cartan matrices, as actually can happen, 
we can always apply some 
simetrization process as it can be done with Kac-Moody algebras, see for example Ref.\cite{cornwell} for a detailed explanation). 
We expect that a  simple root system
$\Delta^0=\{\alpha_1,\ldots ,\alpha_r \}$ and an extended root
system by $\hat \Delta^0={\alpha_0,\alpha_1,\ldots ,\alpha_r }$,
can be constructed. The defining relation is that the (scaled) 
inner product of the roots is 
\begin{eqnarray}
\alpha_i \cdot \alpha_j &=&\hat {\mathbb B}_{ij}
\qquad 1\leq i,j \leq n.
\end{eqnarray}
The set of roots $\alpha_i$ 
are  the simple roots upon which our generalized Cartan Matrix is based. 
They are supposed to play the analogue of a root basis 
of a semisimple Lie Algebra or of a Kac-Moody algebra.
Note that, as happens in KMA Cartan matrices, 
for having the linearly independent 
set of $\alpha_i$ vectors, we generically  
define them in, at least, a $2n-r$ dimensional space $H$. In our case,
as $r=n-1$, we would need a $n+1$ dimensional space. Therefore, the set 
of $n$ roots satisfying the conditions above has to be completed by 
some additional vector, the ``null root'', to obtain a basis for $H$.
The consideration of these complete set of roots will appear in detail 
elsewhere \cite{newtorrente}.

A generic root, $\alpha$, has the form 
$$\alpha=\sum_i c_i \alpha_i$$
where the set of the coefficients $c_i$ are either all non-negative 
integers or all non-positive integers.
In this $n+1$ dimensional space $H$, generic roots can be defined and the same generalized  definition 
for the inner product of two generic roots $\alpha,\beta$ as in 
affine Kac-Moody algebras applies. This generalized definition 
reduces to the inner product above for any two simple roots.

The roots can be of  the ``same length'' or not. A difference with 
respect KMA is that the 
condition of having all roots the same length, symmetry of the matrix 
and of being ``simple laced'' are not equivalent now. This is a simple 
consequence of allowing more than one type of diagonal entries in the 
matrix.

Since $B$ is of rank $r=n-1$, we can find one, and only one, 
non zero vector $\mu$ such that 
$$B\mu=0.$$
The numbers, $ a_i$, components of the vector $\mu$, 
are called Coxeter labels.
The sums of the Coxeter labels  $h=\sum \mu_i$ is the 
Coxeter number.
For a symmetric generalized Cartan matrix only this type of 
Coxeter number appear.

\section{Some simple examples: from graphs to roots}


Let consider the reflexive polyhedron, which corresponds to a 
K3-sliced $CY_3$
space and which is defined by two extended vectors \cite{Vol}
${\vec k}_L^{ext}, {\vec k}_R^{ext}$, one coming from 
the following set of vectors
$S_L=\{(0,0,0,0,{\vec k }_1),(0,0,0,{\vec k }_2), (0,0,{\vec k }_3),..(perms)..\}$,
where the remaining dots correspond to permutations of the position of 
zeroes and vectors $k$, for example permutations of the 
type $\{ (0,k,0,0,0), (k,0,0,0,0)$, etc \}.
The other one comes from the set
$S_R= \{(0,{\vec k }_4), ...(perms)..\} $,  
The vectors ${\vec k}_1,{\vec k}_2,{\vec k}_3$
are respectively any of the five   RWVs of dimension 1,2 and 3. 
The  vector ${\vec k}_4$
correspond to any of the  95 $K3$ RWVs of dimension four.

Let consider, as a simple example, a quintic CY3 which 
 its reflexive polyhedron which is defined by two extended vectors,
$\vec k_{1L}^{(ext)}=(1,0,0,0,0)$ 
and $\vec k_{2R}^{(ext)}=(0,1,1,1,1)$ 
(which correspond to the choice $\vec k_4=(1,1,1,1)$). 

Note that the global slice structure of the $CY_3$ corresponding
to the weight vector $\vec k_5\equiv \vec k^{(ext)}=\veckext_{1L}+\veckext_{2R}=(1,1,1,1,1)$ is determined by {\em both} extended vectors appearing in its 
sum. 

The top and bottom, or left and right, skeletons
of reflexive polyhedron  are  determined by extended  vectors,
$\vec k_{1L}^{(ext)},\vec k_{2R}^{(ext)}$ respectively. 
The left skeleton 
will be a tetrahedron with 4-vertices, 6 edges and a number of internal
points over the edges as indicated in the Figure 3 of Ref.\cite{Vol}.

In the next 
sections, we will consider in turn any of the cases 
corresponding to the vectors 
of different dimension enumerated above.
We will see how everycase will give us a rich variety of 
algebraic generalizations of the Cartan matrices and root systems 
appearing in Kac-Moody algebras.

\subsection{Graph from vector (1)+(1111): Construction of the root systems}

The graph associated to vector $\veck_L=(1)$, (or $(10000)$,including all zero 
components), and assuming  $\veck_R= (1111)$ is a closed graph, a tetrahedron (see Ref.\cite{Vol} for 
other type of graphs when combining with other $\veck_R$ vectors). 
Following the rules given above we can 
form the associated Berger matrix, or generalized Cartan matrix.
This matrix is a $28\times 28 $ matrix corresponding to four vertices and
24 internal nodes over the six edges. One can actually check that this matrix 
is of affine type. 
Inspired by this graph we define a full family of graphs where the 
number of internal nodes over the edges are altered. Let us focus in 
the most simple case, where the all internal nodes are eliminated and 
we are left with only the four vertices (figure \ref{fig1}). From this simplified 
graph, not directly obtained from the CY polyhedra, we built the 
corresponding matrix according to the rules given in the previous 
section, we will call this a  $ \CYB_3^{(1)}$ matrix.
Let us  also consider for the sake of comparison 
the graphs and Cartan matrices for the 
affine algebra $A_2^{(1)}$. The associated graph, a standard 
Coxeter-Dynkin diagram, is just one of the triangles corresponding 
to the faces of this simple tetrahedron.
The well known Cartan matrix for the $A_2^{(1)}$ affine algebra and 
our  Berger matrix $ CY3B_3^{(1)}$ are respectively.

\begin{eqnarray}
A_2^{(1)}=
\left (
\begin{array}{ccc}
 2     &-1   & -1  \\
-1     & 2   & -1  \\
-1     &-1   &  2  
\end{array}
\right ),
\qquad 
CY3{\mathbb B}_3^{(1)}=
\left (
\begin{array}{cccc}
 3     &-1   & -1  & -1  \\
-1     & 3   & -1  & -1  \\
-1     &-1   &  3  & -1  \\
-1     &-1   & -1  &  3  
\end{array}
\right )
\end{eqnarray}

\FIGURE{

{ \unitlength1.3mm
\centering
\begin{tabular}{rcl} 

\hspace{1cm}
  \begin{picture}(50,50)
   \multiput(0,0)(10,0){2}{\circle*{1.7}}
   \put(0,0){\line(1,0){10}}
   \multiput(1,2)(10,0){2}{1}
   \end{picture}  
\hspace{-3cm}
&

   \begin{picture}(50,50)
   \multiput(0,0)(20,0){2}{\circle*{1.7}}
   \put(0,0){\line(1,0){20}}
   \multiput(10,10)(10,0){1}{\circle*{1.7}}
   \put(0,0){\line(1,1){10}}
   \put(10,10){\line(1,-1){10}}
   \multiput(1,2)(20,0){2}{1}
   \multiput(11,12)(10,0){1}{1}
   \end{picture}  

\hspace{-2cm}
&

   \begin{picture}(50,50)
   \multiput(0,0)(20,0){2}{\circle*{1.7}}
   \put(0,0){\line(1,0){20}}
   \multiput(12,10)(10,0){1}{\circle*{1.7}}
   \multiput(12,-10)(10,0){1}{\circle*{1.7}}
   \multiput(0,0)(20,0){2}{\circle{3}}
   \multiput(12,10)(10,0){1}{\circle{3}}
   \multiput(12,-10)(10,0){1}{\circle{3}}
   \put(12,-10){\line(0,1){20}}
   \put(0,0){\line(6,5){11.1}}
   \put(0,0){\line(6,-5){11.1}}
   \put(12,10){\line(4,-5){8.1}}
   \put(12,-10){\line(4,5){8.1}}
   \multiput(1,2)(20,0){2}{1}
   \multiput(13,12)(10,0){1}{1}
   \multiput(13,-8)(10,0){1}{1}
   \end{picture}  
\end{tabular}
}
\vspace{1.3cm}
\label{fig1}
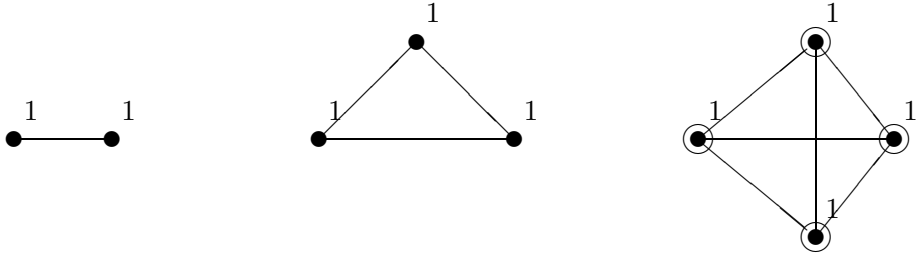
\caption{Respectively from left to right, 
the Dynkin diagrams of $A_1,A_2^{(1)}$ and the generalized 
Berger-Dynking diagram tetrahedron dyagram $CY3B_3$. 
Coxeter Labels are attached at the nodes.
Any graph of the series is an edge or face of the following 
graph. 
In this and following graphs double-circled nodes yield 
diagonal entries equal to three in the corresponding matrix.
}
}

As the next step, we build a system of 
vectors, the equivalent of a root system,
 which satisfy the Berger matrix interpreted as an inner product.
Let us remind first that 
for the well known affine case $A_2^{(1)} (n=3,r=1)$,  
supposing a set of orthonormal canonical basis 
$\{ e_i \}$, one  obtains  
 three vectors  $\alpha_i$ such as:
\begin{eqnarray}
\alpha_1&=&e_1-e_2, \nonumber\\
\alpha_2&=&e_2-e_3 ,\nonumber\\
\alpha_3&=&e_3-e_1, \nonumber
\end{eqnarray}
the root 
$\alpha_0 \equiv -\alpha_3$ is the eldest root vector in the 
$A_2$ algebra.
For the Cartan matrix of  
the affine Kac-Moody algebra,  $A_2^{(1)}$, the sum of 
all simple roots with certain coefficients, the Coxeter labels,
 is equal zero. Diagonalizing and obtaining the zero mode of the matrix,
one can easily see that 
for the  full  $A_r^{(1)}$ series  all  the Coxeter labels are equal to one 
(see for example \cite{FSS,CSM}), we have:
\begin{eqnarray}
1 \cdot \alpha_1+ 1 \cdot \alpha_2+1 \cdot \alpha_3 = 0.
\end{eqnarray}
For the case corresponding to  $CY{\mathbb B}_3^{(1)}$, 
a symmetric $n=4,r=3$ matrix,   
we obtain the following simple root system. 
As we mentioned before we 
 need a vector space of dimension $d>2n-r=5$ to express the roots, 
we use in this example an overcomplete 
6 dimensional orthonormal basis of vectors $(\{e_i\},i=1,...,6)$. 
The advantage of this basis, among many other choices, being that 
the coefficients of the vectors are specially simple here:
\begin{eqnarray}
\alpha_1&=&e_1+e_2+e_3, \nonumber\\
\alpha_2&=&-e_1-e_4+e_5 \nonumber\\
\alpha_3&=&-e_2+e_4+e_6, \nonumber\\
\alpha_4&=&-e_3-e_5-e_6, \nonumber
\end{eqnarray}

It is easy to check that these ¨simple roots¨ 
satisfy   all the  restrictions imposed by the Berger matrix
$\alpha_{i} \cdot \alpha_{j}=B_{ij}$.
Following the nomenclature of Kac-Moody algebras we say that 
all roots in this case are of the same length: a property of being 
``simple laced''. We will see however that the concepts of 
simply laced, symmetric matrices and equal-length roots are not 
totally equivalent for Berger matrices and should be distinguished.

The diagonalization of the matrix gives us the zero mode vector, 
$B\mu=0$. In this case $\mu=(1,1,1,1)$ and $h=4$. 
The affine condition satisfied by the 
set of simple roots:
\begin{eqnarray}
1 \cdot \alpha_1+ 1 \cdot \alpha_2+1 \cdot \alpha_3+1\cdot \alpha_4 = 0,
\end{eqnarray}
the coefficients of this vector being the Coxeter labels, components 
of the vector $\mu$.

The Berger matrix as $CY3B3$ and its tetrahedron 
graph can be easily generalized to any dimension. We can consider 
an infinite series of matrices and hyper-tetrahedron graphs originated 
from $CY_n$ spaces defined as those $(n+1)\times (n+1)$ matrices  with all non-diagonal entries 
equal  to $-1$ and $n$ in the diagonal.
For illustration, the matrix corresponding to 
$CY4{\mathbb B}_4^{(1)}$ is 
 \begin{eqnarray}
CY4{\mathbb B}_4^{(1)}=
\left (
\begin{array}{ccccc}
 4     &-1   & -1  & -1 & -1  \\
-1     & 4   & -1  & -1 & -1  \\
-1     &-1   &  4  & -1 & -1  \\
-1     &-1   & -1  &  4 & -1  \\
-1     &-1   & -1  & -1 &  4  
\end{array}
\right ),
\end{eqnarray}
the determinant of this matrix is effectively zero
 $\det(CY4{\mathbb B}_4^{(1)})=0$. 

For any dimension $n$ 
it is straightforward to 
show that $\det (\mathbf{-1}+x\ I)=(x-n) x^{n-1}$ 
where $\mathbf{1},I$ are respectively the matrix filled with 
$1$ and the indentity matrix. 
Thus, if we choose  for the diagonal entries the value
 $x=n$, the full 
family of matrices $CYnB$ adquires zero determinant. 
According to the previous formula for the determinant, this choice 
is {\em unique}. It is also straightforward to show that for the
full family of matrices we have the vector of Coxeter labels
$\mu_{(n)}=(1,..(n)..1)$ and the affine condition 
$\sum_n \alpha_i=0$.

\subsubsection{Additional algebraically defined Berger matrices of any order.}

In the previous paragraphs, we have used the definition of the 
Berger graphs and matrices 
independently from their Calabi-Yau polyhedra origin, 
we want to extend this single matrix to a infinite series of 
related matrices.
On purely algebraic grounds we consider a  series of new Berger graphs and
matrices that we will denote as  $CY3B_s^{(1)}, s\geq 3$.
The graphs are obtained  including  any number 
 of  internal  Cartan nodes on one or more than one  of  the edges of the 
defining tetrahedron considered in the previous section. 
An arbitrary example of a tetrahedron graph with a number of internal nodes
is given in Fig.(\ref{fig2}).  We consider this graph as a generalization 
of affine Kac-Moody algebras:
each face of the tetrahedron resembles a $A_r^{(1)}$ triangle graph, 
where $r+1$ internal nodes appear. 

From this enlarged graph 
we build corresponding Berger matrices using the rules 
given previously. 
These matrices will differ 
from the basic 
$\CYB_3^{(1)}$ matrix above by the appearance 
of $s-3$ elements equal to two in 
the diagonal, corresponding to the new internal nodes. The diagonal 
elements corresponding to the unchanged vertex nodes remain the same.

The new matrices thus built are ``well-behaved'' Berger matrices: 
The algebraic Berger conditions are fulfilled for any number of edge nodes.
Thus we have an infinite series of matrices in parallel
to what happens in the Cartan-Lie case with the series $A_r^{(1)}$.
Let us note however that for this extension, no reference to Calabi-Yau 
geometry has been used.

As an illustrating example, let us write thee  cases where $s=4,5$, we include 
one or two new internal nodes respectively 
in one of the edges of the tetrahedron. The  the  Berger matrices
are
\begin{eqnarray}
CY3{\mathbb B}_4^{(1)}=
\left (
\begin{array}{ccccc}
   2   & -1  &  -1 &  0 &   0  \\
  -1   &  3  &   0 & -1 &  -1 \\
  -1   &  0  &   3 & -1 &  -1 \\
   0   &  0  &  -1 &  3 &  -1 \\
   0   &  0  &  -1 & -1 &   3 
\end{array}
\right ),\quad
&
CY3{\mathbb B}_5^{(1)}=
\left (
\begin{array}{cccccc}
 3     & -1   &  0  &  0 &  -1 & -1  \\
-1     &  2   & -1  &  0 &   0 &  0 \\
 0     & -1   &  2  & -1 &   0 &  0 \\
 0     &  0   & -1  &  3 &  -1 & -1 \\
-1     &  0   &  0  & -1 &   3 & -1 \\
-1     &  0   &  0  & -1 &  -1 &  3 
\end{array}
\right ).
\end{eqnarray}

One can compute explicitly  that  the determinant of these matrices is  
equal to 
zero and that all the principal proper minors are positive. Clearly 
the position of the diagonal elements is unimportant and can be 
changed by relabeling of the nodes.

We proceed now to obtain  a system of simple roots, we show 
just as illustration the case corresponding to 
 the second matrix, $CY3{\mathbb B}_5^{(1)}$. 
One needs in this case
  a vector space of dimension $d\geq 2n-r=7$, where $r=1$, to express the roots, 
we use again an overcomplete 
set  ($\{e_i\},i=1,...,8$ where the 
coefficients appears specially simple:
\begin{eqnarray}
\alpha_1     &=& e_1+e_2+e_3, \nonumber\\
\alpha_2     &=& -e_1+e_7,     \nonumber\\
\alpha_3     &=& -e_7+e_8      \nonumber\\
\alpha_4     &=& -e_8-e_4+e_5, \nonumber\\
\alpha_5     &=& -e_2+e_4+e_6, \nonumber\\
\alpha_6     &=& -e_3-e_5-e_6. \nonumber
\end{eqnarray}
The root system  contains now  
two additional binary  roots $\alpha_2,\alpha_3$.
The roots are not of the ``same length'': as we have mentioned before the 
condition of having all roots the same length, symmetry of the matrix 
and of being ``simple laced'' are not equivalent now. This is a simple 
consequence of allowing more than one type of diagonal entries in the 
matrix.
The Coxeter labels and affine condition on the roots are easily obtained.
The diagonalization of the matrix gives us the zero mode vector, 
$B\mu=0$. In this case $\mu=(1,1,1,1,1,1)$ and $h=6$. 
The affine condition satisfied by the 
set of simple roots:
$\sum_6 \alpha_6=0$
the coefficients of this vector being the Coxeter labels, components 
of the vector $\mu$.

\FIGURE{{ \unitlength1.3mm
\centering
\begin{tabular}{lr}
\hspace{2cm} 
   \begin{picture}(50,50)
   \multiput(0,0)(20,0){2}{\circle*{1.7}}
   \put(0,0){\line(1,0){20}}
   \multiput(12,10)(10,0){1}{\circle*{1.7}}
   \multiput(12,-10)(10,0){1}{\circle*{1.7}}
   \multiput(0,0)(20,0){2}{\circle{3}}
   \multiput(12,10)(10,0){1}{\circle{3}}
   \multiput(12,-10)(10,0){1}{\circle{3}}
   \multiput(12,-3.333)(10,0){1}{\circle*{1.7}}

   \put(12,-10){\line(0,1){20}}
   \put(0,0){\line(6,5){11.1}}
   \put(0,0){\line(6,-5){11.1}}
   \put(12,10){\line(4,-5){8.1}}
   \put(12,-10){\line(4,5){8.1}}

   \multiput(1,2)(20,0){2}{1}
   \multiput(13,12)(10,0){1}{1}
   \multiput(13,-8)(10,0){1}{1}
   \multiput(13,-1.333)(10,0){1}{1}
   \end{picture}

&
\hspace{-1cm}
   \begin{picture}(50,50)
   \multiput(0,0)(20,0){2}{\circle*{1.7}}
   \put(0,0){\line(1,0){20}}
   \multiput(12,10)(10,0){1}{\circle*{1.7}}
   \multiput(12,-10)(10,0){1}{\circle*{1.7}}
   \multiput(0,0)(20,0){2}{\circle{3}}
   \multiput(12,10)(10,0){1}{\circle{3}}
   \multiput(12,-10)(10,0){1}{\circle{3}}
   \multiput(12,3.333)(10,0){1}{\circle*{1.7}}
   \multiput(12,-3.333)(10,0){1}{\circle*{1.7}}

   \put(12,-10){\line(0,1){20}}
   \put(0,0){\line(6,5){11.1}}
   \put(0,0){\line(6,-5){11.1}}
   \put(12,10){\line(4,-5){8.1}}
   \put(12,-10){\line(4,5){8.1}}

   \multiput(1,2)(20,0){2}{1}
   \multiput(13,12)(10,0){1}{1}
   \multiput(13,-8)(10,0){1}{1}
   \multiput(13,5.333)(10,0){1}{1}
   \multiput(13,-1.333)(10,0){1}{1}
   \end{picture}  

\end{tabular}}\vspace{1.3cm}
\label{fig2}
\caption{
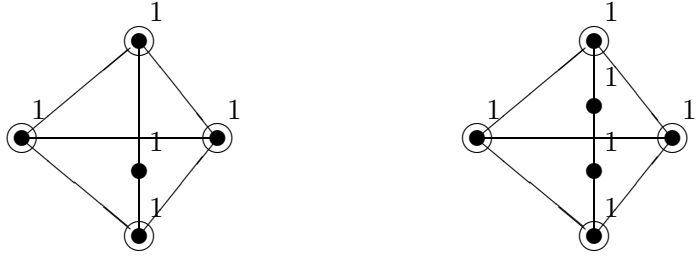
Generalized Berger-Dynking  
tetrahedron dyagram $CY3B_3$ including  one or two additional 
internal nodes in one of the edges. }
}

\subsection{The left (111)-right (1111)  case.}

The graph associated to the case where we have a  
left vector $(111)$ plus a right vector (1111) as before, can equally 
be extracted from the reflexive Newton polyhedron 
following a simple procedure.
The result appears in Fig.(\ref{fig3}). In this case, we directly obtain the 
Coxeter-Dynkin diagram corresponding to the affine algebra $E_6^{(1)}$. 
We can easily check that 
 following the rules given above we can 
form an associated Berger matrix, which, coincides with 
the  corresponding  generalized Cartan matrix of the 
the affine algebra $E_6^{(1)}$. 
The well known Cartan matrix for this is:
\begin{eqnarray}
E_6^{(1)}=CY B3&=&
\pmatrix{ 2&-1&  0& 0&   0& 0&   0\cr
-1& 2&  0& 0&   0& 0&  -1\cr
 0& 0&  2&-1&   0& 0&   0\cr
 0& 0& -1& 2&   0& 0&  -1\cr
 0& 0&  0& 0&   2&-1&   0\cr
 0& 0&  0& 0&  -1& 2&  -1\cr
 0&-1&  0&-1&   0&-1&   2}
\end{eqnarray}

The root system is well known, we have 
(in a, minimal, ortonormal basis $(\{e_i\},i=1,...,8$):
\begin{eqnarray}
\alpha_1   &=& -\frac{1}{2}(-e_1+e_2+e_3+e_4+e_5+e_6+e_7-e_8)\nonumber\\      
\alpha_2   &=&      (e_2-e_1)                             \nonumber\\
\alpha_3   &=&      (e_4-e_3)                             \nonumber\\
\alpha_4   &=&      (e_5-e_4)                             \nonumber\\
\alpha_5   &=&      (e_1+e_2)                             \nonumber\\
\alpha_6   &=& -\frac{1}{2}(e_1+e_2+e_3+e_4+e_5-e_6-e_7+e_8)  \nonumber  \\   
\alpha_7   &=& -e_2+e_3                          \nonumber  
\end{eqnarray}
Coxeter labels and affine condition are easily reobtained.
The diagonalization of the matrix gives us the zero mode vector, 
$B\mu=0$. In this case the Coxeter labels are $\mu=(1,2,1,2,1,2,3)$
 and $h=12$. 
The affine condition satisfied by the 
set of simple roots is also well known 
$$\alpha_1+2 \alpha_2+\alpha_3+2\alpha_4+\alpha_5+2 \alpha_6+3\alpha_7=0$$.

\FIGURE{{ \unitlength0.8mm
\centering
\begin{tabular}{rr} 
   \begin{picture}(50,50)
   \multiput(0,0)(10,0){5}{\circle*{1.7}}
   \put(0,0){\line(1,0){40}}
   \multiput(20,0)(0,10){2}{\circle*{1.7}}
   \put(20,0){\line(0,1){10}}

    \put(1,2){1}
    \put(11,2){2}
    \put(21,2){3}
    \put(31,2){2}
    \put(41,2){1}
    \put(21,2){3}
    \put(21,12){2}
   \end{picture}

&

   \begin{picture}(50,50)
   \multiput(0,0)(10,0){5}{\circle*{1.7}}
   \put(0,0){\line(1,0){40}}
   \multiput(20,0)(0,10){3}{\circle*{1.7}}
   \put(20,0){\line(0,1){20}}
   
    \put(1,2){1}
    \put(11,2){2}
    \put(21,2){3}
    \put(31,2){2}
    \put(41,2){1}
    \put(21,2){3}
    \put(21,12){2}
    \put(21,22){1}

   \end{picture}  

\end{tabular}}\vspace{1.cm}
\label{fig3}\caption{
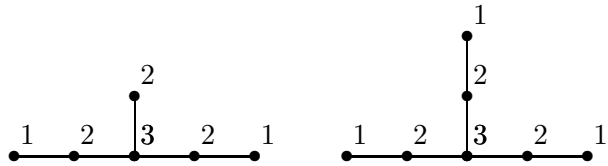
Dynkin diagrams for the exceptional $E_6$ and affine $E_6^{(1)}$ algebras 
(left and right respectively).}
}

Now, we ask the question on how we can generalize the appearance 
this affine algebra using our Berger construction. 
We can consider at least two different ways.

In the first way we will consider the standard graph and matrix corresponding
to $E_6^{(1)}$ and attach an increasing number of legs to the central 
node. The number of nodes in each leg  is a well defined number: equals 
the number of legs itself minus one. For example in the $E_6^{(1)}$ case,
associated to vector $(111)[3]$ 
we have three legs with two nodes in each leg, successive diagrams will
include 4,5.. legs with 3,4.. nodes. Loosely speaking, we say that 
these Dynkin graphs are associated to the series of vectors
 $(1,...,1)[n]$ for any $n$. 
In the second way, we will start again from 
 the standard graph and matrix corresponding
to $E_6^{(1)}$ and insert additional internal nodes. 
Let us note that one can show that
  internal nodes can be added or appended to the legs of the graph, 
this would be already known extensions
of the affine algebra (see for example,Ref.\cite{west2002}). In our context, this kind 
of extension would be just an example of the addition of internal nodes as we 
did in the previous section, similarly we could see that we can draw graphs 
and obtain matrices in a consistent way.  

But our procedure allow us for more complicated extensions: we can add
not only just nodes to the legs, but legs itselfs.
 In this way we are able to 
glue together in a non-trivial way Coxeter-Dynkin diagrams corresponding 
to different affine algebras and obtain Berger(-Coxeter-Dynkin) matrices 
consistent with the definitions given above. 

Surprisingly, these two ways of generalization seem to be mutually 
exclusive. No example is known to us where both can be simultaneously 
performed (see discussion at the end of section 3.3). 

As we have mentioned before,
we somehow generalize the decomposition of Kac-Moody algebras. Theory of 
Kac-Moody algebras show us that for any {\em finite} or {\em affine} Kac-Moody
 algebra, every proper subdiagram 
(defined as that part of the 
generalized Coxeter-Dynkin diagram obtained by removing one or more vertices 
and the lines attached to these vertices) 
is a collection of diagrams corresponding to {\em finite} Kac-Moody algebras.
In our case we have more flexibility. Proper subdiagrams, obtained 
eliminating internal nodes or vertices, are in general collections of 
Berger-Coxeter-Dynkin diagrams corresponding to other (affine by construction )Berger diagrams 
{\em or} to {\em affine} Kac-Moody algebras.
This property clearly paves the way to the consideration of non-trivial 
extensions of SM and string symmetries.

In the next two sections, examples of each of the two procedure outlined 
above will be treated in turn.

\subsection{An ``exceptional'' series of graphs associated to  $(1,...,1)$}

We  discuss the first way of generalization  explained above.
One takes the standard graph and matrix corresponding
to $E_6^{(1)}$ and attaches an increasing number of legs to the central 
node. The number of nodes in each leg  is a well defined number: equals 
the number of legs itself minus one. For example in the $E_6^{(1)}$ case,
associated to vector $(111)[3]$ 
we have three legs with two nodes in each leg, successive diagrams will
include 4,5.. legs with 3,4.. nodes. Loosely speaking, we say that 
these Dynkin graphs are associated to the series of vectors
 $(1,...,1)[n]$ for any $n$.  
The graph associated to vector $(1111)$  appears 
in Fig.\ref{fig4} (center). This one have  much more  different structure comparing to the previous
Berger graphs. Its more important characteristic  it is that it
 has a vertex-node with outgoing 4-lines.  
We call this graph a generalized $E_6^{(1)}$ graph of type I.

The Berger matrix is obtained from the planar graph according to the standard
rules.  We assign different values (2 or 3 ) to diagonal entries depending 
if they are associated to standard nodes or to the central vertex.
The result is the following $13\times 13$ symmetric matrix 
containing, as more significant difference, an additional 3 diagonal entry:
{\small
\begin{eqnarray}
CY3B_{()}&=&\left (
\begin{array}{ccc|ccc|ccc|ccc|c}
 2 &-1 & 0   & 0 & 0 & 0   & 0 & 0 & 0   & 0 & 0 & 0  &-1 \\
-1 & 2 &-1   & 0 & 0 & 0   & 0 & 0 & 0   & 0 & 0 & 0  & 0 \\
 0 & -1& 2   & 0 & 0 & 0   & 0 & 0 & 0   & 0 & 0 & 0  & 0 \\
\hline
 0 & 0 & 0  & 2 &-1 & 0    & 0 & 0 & 0   & 0 & 0 & 0  &-1 \\
 0 & 0 & 0  &-1 & 2 &-1    & 0 & 0 & 0   & 0 & 0 & 0  & 0 \\
 0 & 0 & 0  & 0 &-1 & 2    & 0 & 0 & 0   & 0 & 0 & 0  & 0 \\
\hline
 0 & 0 & 0  & 0 & 0 & 0    & 2 &-1 & 0   & 0 & 0 & 0  &-1 \\
 0 & 0 & 0  & 0 & 0 & 0    &-1 & 2 &-1   & 0 & 0 & 0  & 0 \\
 0 & 0 & 0  & 0 & 0 & 0    & 0 & 1 & 2   & 0 & 0 & 0  & 0 \\
\hline
 0 & 0 & 0  & 0 & 0 & 0    & 0 & 0 & 0   & 2 &-1 & 0  &-1 \\
 0 & 0 & 0  & 0 & 0 & 0    & 0 & 0 & 0   &-1 & 2 &-1  & 0 \\
 0 & 0 & 0  & 0 & 0 & 0    & 0 & 0 & 0   & 0 &-1 & 2  & 0 \\
\hline
-1 & 0 & 0  &-1 & 0 & 0    &-1 & 0 & 0   &-1 & 0 & 0  & 3 
\end{array}
\right )
\end{eqnarray}
}
One can check that this matrix fulfills the conditions for 
Berger matrices. Its determinant is zero while the rank $r=12$. 
All the principal minors are positive.

One can obtain a system of roots $(\alpha_i, i=1,\ldots,13 )$ 
in a orthonormal basis.
Considering the  orthonormal canonical basis 
$(\{e_i\}, i=1,\ldots, 12)$, we obtain:

\begin{eqnarray}
\alpha_1&=& -(e_1-e_2)                                            \nonumber\\
\alpha_2&=& \frac{1}{2} [(e_1 - e_2 - e_3+e_4+e_5+e_6)+(e_8-e_7)]  \nonumber\\
\alpha_3&=&-(e_8- e_7)  \nonumber\\
\alpha_4&=&(e_4- e_3)  \nonumber\\
\alpha_5&=&(e_5- e_4)  \nonumber\\
\alpha_6&=&(e_6- e_5)  \nonumber\\
\alpha_7&=&(e_1+ e_2)  \nonumber\\
\alpha_8&=&   -\frac{1}{2} [(e_1 + e_2 - e_9-e_{10}-e_{11}-e_{12})+(e_8+e_7)]
\nonumber\\
\alpha_9&=& (e_8+e_7)
 \nonumber\\
\alpha_{10}&=& -(e_{10}-e_9) \nonumber\\
\alpha_{11}&=& -(e_{11}-e_{10}) \nonumber\\
\alpha_{12}&=& -(e_{12}-e_{11}) \nonumber\\
\alpha_{13}&=&  e_3-e_2-e_9 \nonumber
\end{eqnarray}

The assignment of roots to the nodes of the Berger-Dynkin graph 
is given in Fig.\ref{fig5}.
It easily to check the inner product of these ¨simple roots¨ 
leads to the Berger Matrix $a_i\cdot a_j=B_{ij}$.
This matrix has one null eigenvector, with coordinates, in the $\alpha$ basis,
$\mu=(3,2,1,3,2,1,3,2,1,3,2,1,4).$
The Coxeter number is $h=22$. One can 
check that these Coxeter labels are identical to those obtained from 
the  geometrical construction \cite{CF,Vol}. They are shown explicitly in 
Fig.\ref{fig4}.
Correspondingly 
the following linear combination of the ¨roots¨ satisfies the affine 
condition: 
\begin{eqnarray}
4 \alpha_{0}+
3 \alpha_{1}     + 2 \alpha_{2}  +     \alpha_{3}  +  
3 \alpha_{4}     + 2 \alpha_{5}  +     \alpha_{6}  +  
3 \alpha_{7}     + 2 \alpha_{8}  +     \alpha_{9}  + 
3 \alpha_{10}    + 2 \alpha_{11} +    \alpha_{12} & =& 0 \nonumber 
 \end{eqnarray}

\FIGURE{
\parbox{\linewidth}{\centering
 \unitlength0.7mm
\begin{tabular}{lrr} 

   \begin{picture}(50,50)
   \multiput(0,0)(10,0){5}{\circle*{1.7}}
   \put(0,0){\line(1,0){40}}
   \multiput(20,0)(0,10){3}{\circle*{1.7}}
   \put(20,0){\line(0,1){20}}
   
    \put(1,2){1}
    \put(11,2){2}
    \put(21,2){3}
    \put(31,2){2}
    \put(41,2){1}
    \put(21,2){3}
    \put(21,12){2}
    \put(21,22){1}

   \end{picture}

&

\hspace{0.1cm}

   \begin{picture}(50,50)
   \multiput(0,0)(10,0){7}{\circle*{1.7}}
   \put(0,0){\line(1,0){60}}
   \multiput(30,-30)(0,10){7}{\circle*{1.7}}
   \put(30,-30){\line(0,1){60}}
    \put(30,0){\circle{3}}
    \put(1,2){1}
    \put(11,2){2}
    \put(21,2){3}
    \put(31,2){4}
    \put(41,2){3}
    \put(51,2){2}
    \put(61,2){1}
    \put(31,-28){1}
    \put(31,-18){2}
    \put(31,-8){3}
    \put(31,2){4}
    \put(31,12){3}
    \put(31,22){2}
    \put(31,32){1}
   \end{picture}

&

\hspace{1.5cm}
   \begin{picture}(50,50)
   \multiput(0,0)(10,0){9}{\circle*{1.7}}
   \put(0,0){\line(1,0){80}}

   \multiput(40,0)(0,10){5}{\circle*{1.7}}
   \put(40,0){\line(0,1){40}}

   \multiput(40,0)(-4.47,-8.94){5}{\circle*{1.7}}
   \put(40,0){\line(-1,-2){16.775}}

   \multiput(40,0)(4.47,-8.94){5}{\circle*{1.7}}
   \put(40,0){\line(1,-2){16.775}}


    \put(40,0){\circle{3}}
    \put(40,0){\circle{4}}

    \put(1,2){1}
    \put(11,2){2}
    \put(21,2){3}
    \put(31,2){4}
    \put(41,2){5}
    \put(51,2){4}
    \put(61,2){3}
    \put(71,2){2}
    \put(81,2){1}
    \put(41,12){4}
    \put(41,22){3}
    \put(41,32){2}
    \put(41,42){1}
\newcounter{cms}
\setcounter{cms}{6}
   \multiput(40,0)(-4.57,-8.94){5}{\addtocounter{cms}{-1}\small\arabic{cms}}
\setcounter{cms}{6}
   \multiput(40,0)(4.57,-8.94){5}{\addtocounter{cms}{-1}\small\arabic{cms}}
   \end{picture}  

\end{tabular}}\vspace{2.4cm}
\label{fig4}
\caption{
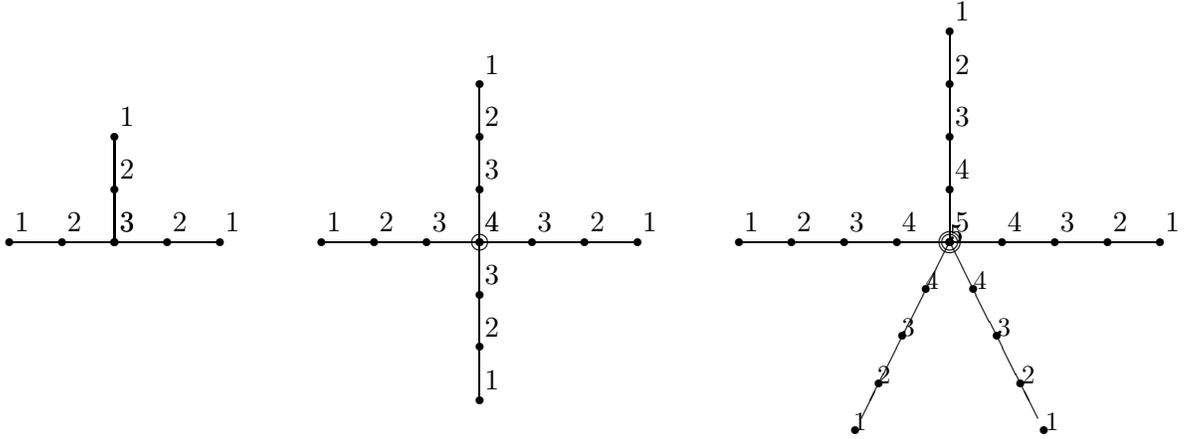
Berger-Dynkin diagrams for the 
 affine $E_6^{(1)}$ and its generalizations 
$CY3-E_6^{(1)}$ and  $CY4-E_6^{(1)}$ diagrams  (left and right respectively).}
}

One could try to pursue the generalization process of graphs 
and matrices adding internal nodes to this case as it has been done 
previously. Surprisingly, in contradiction to previous case where 
an infinite series of new graphs and matrices can be obtained, this is 
however and ``exceptional'' case. No infinite series of graphs can be 
obtained in this way.

However, we can suppose that the graphs 
 corresponding to the vectors $(111), (1111),(1,..n..,1)$ 
(see Fig.\ref{fig4}), 
generate  an infinite series 
of new Cartan Matrices 
$CY_{2,3,4,...}E_6^{(1)}$ ordered  with respect to Calabi-Yau 
space dimension.
 The first term in this infinite series 
starting from the $E_6^{(1)}$ case.
All our three restrictions to the Berger graphs can be easily checked.
The determinant of these matrices for all cases is equal zero.

Similarly, one can easily find \cite{Vol} the dimensional generalizations of 
the $E_7^{[1]}$
and  $E_8^{[1]}$ graphs (corresponding to the choice of 
vectors (112),(1123)), ``exceptional'' cases themselves in the same 
sense, which give us the infinite series linked with
the  dimension of $d$. A complete exposition of these cases will appear 
in Ref.\cite{newtorrente}.

Let us finish this illustratory example mentioning the possibility 
of recovering the Coxeter-Dynkin diagrams products of finite 
Lie algebras or affine Kac-Moody algebras by eliminating one the 
roots appearing above. It is apparent from the example shown in 
Fig.\ref{fig6} that they reappear as proper subdiagrams when 
removing one or more vertices and the lines attached to these 
vertices.

\FIGURE{
\parbox{\linewidth}{
\centering
 \unitlength1.4mm
\hspace{-1cm}
   \begin{picture}(50,50)
   \multiput(0,0)(10,0){7}{\circle*{1.7}}
   \put(0,0){\line(1,0){60}}
   \multiput(30,-30)(0,10){7}{\circle*{1.7}}
   \put(30,-30){\line(0,1){60}}
    \put(30,0){\circle{3}}
    \put(0,-5){\small $e_6-e_5$}
    \put(10,-5){\small $e_5-e_4$}
    \put(20,-5){\small $e_4-e_3$}
    \put(30,-5){\small $$}
    \put(40,-5){\small $e_9-e_{10}$ }
    \put(50,-5){\small $e_{10}-e_{11}$ }
    \put(60,-5){\small $e_{11}-e_{12}$ }
    \put(32,-30){\small$e_7+e_8$}
    \put(32,-20){\small$-1/2 (e_2+e_1-e_9-e_{10}-e_{11}-e_{12}+e_7+e_8)$}
    \put(32,-10){\small$e_2+e_1$}
    \put(32,0){\small  $$}
    \put(32,10){\small $e_2-e_1$}
    \put(32,20){\small$-1/2 (e_2-e_1+e_3+e_4+e_5+e_6+e_7-e_8)$}
    \put(32,30){\small $e_7-e_8$}

    \put(15,5){\small $e_3-e_2-e_9$}
    \put(24,5){\vector(1,-1){4}}
   \end{picture}

}
\hphantom{\hfill}
\vspace{4cm}
\label{fig5}
\caption{
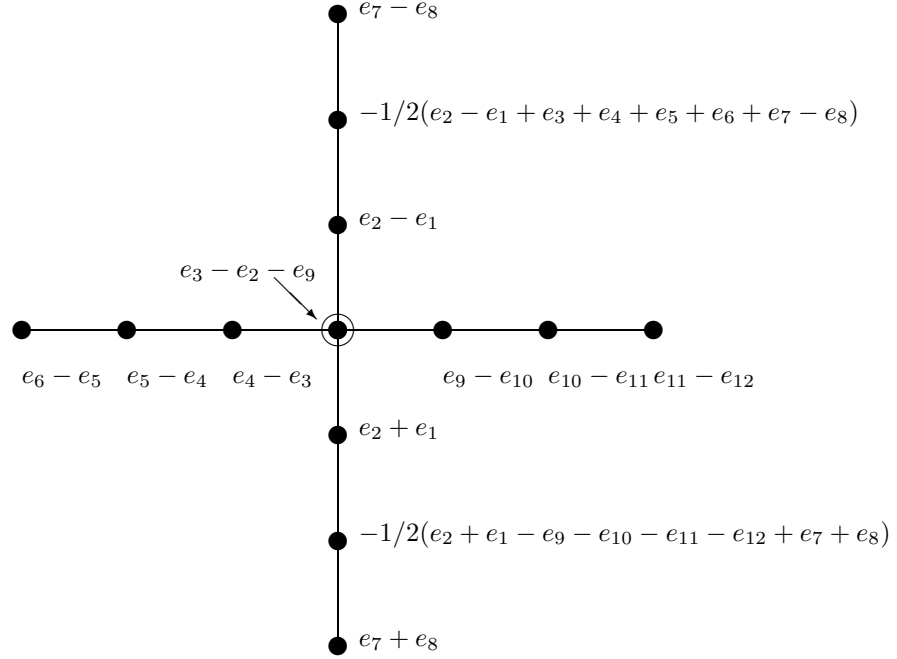
Berger-Dynkin diagram and root system for the 
$CY3-E_6^{(1)}$ matrix.}
}

\subsection{Another generalization: the (111) graph plus internal nodes}

As a new example,
in this section we will follow a different way of generalization 
of the $E_6^{(1)}$ graph
on pure 
algebraic ways, independently of Calabi-Yau classification arguments.
For this purpose, we take as starting point
 the Berger graph $E_6^{(1)}$ 
corresponding to the RWV vector $(111)$, shown in Figs.\ref{fig3} or \ref{fig4},
 and we add a fourth leg to the central node. At the end of the new leg 
we will attach a new copy of a $E_6^{(1)}$ Coxeter-Dynkin diagram. 
The length of this internal leg connecting the two affine algebra copies
 is arbitrary.

The important fact of this seemingly arbitrary construction is that 
if write the corresponding matrix, according to the 
rules Eqs.\ref{eqsberger}, we obtain a new Berger matrix, that is, a 
matrix which fulfills the same defining conditions. 
In particular the  affine condition  and
 the positivity conditions.
This fact lead us again to consider the importance on its own of the 
Berger matrix definition as a generalization of the Kac-Moody Cartan matrix.
Thus we are lead to 
consider the series of Berger graphs and matrices built according 
to this algorithm as infinite: in an obvious 
symbolic notation $E_6^{(1)}+A_r^{(1)}+E_6^{(1)}$ where $r$ is
arbitrary.

As an example we draw in Fig. \ref{fig6} the graph resulting 
from the incorporation of a two node extra leg. The total number of nodes, in 
obvious notation, is   $7_L+2_{int}+7_R=16$. 
The $16 \times 16$
Berger matrix for this graph is the following. 

{\small
\begin{eqnarray}
CY3B&=&\left (
\begin{array}{ccccccc|cc|ccccccc}
 2 &-1 & 0 & 0 & 0 & 0 &  0 &  0 & 0    & 0 & 0 & 0 & 0 & 0 & 0 & 0 \\
-1 & 2 & 0 & 0 & 0 & 0 & -1 &  0 & 0    & 0 & 0 & 0 & 0 & 0 & 0 & 0 \\
 0 & 0 & 2 &-1 & 0 & 0 &  0 &  0 & 0    & 0 & 0 & 0 & 0 & 0 & 0 & 0 \\
 0 & 0 &-1 & 2 & 0 & 0 & -1 &  0 & 0    & 0 & 0 & 0 & 0 & 0 & 0 & 0 \\
 0 & 0 &0  & 0 & 2 &-1 &  0 &  0 & 0    & 0 & 0 & 0 & 0 & 0 & 0 & 0 \\
 0 & 0 &0  & 0 &-1 & 2 & -1 &  0 & 0    & 0 & 0 & 0 & 0 & 0 & 0 & 0 \\
 0 &-1 &0  &-1 & 0 &-1 &  3 & -1 & 0    & 0 & 0 & 0 & 0 & 0 & 0 & 0 \\
\hline
 0 & 0 &0  & 0 & 0 & 0 & -1 &  2 &-1    & 0 & 0 & 0 & 0 & 0 & 0 & 0 \\
 0 & 0 &0  & 0 & 0 & 0 &  0 & -1 & 2    &-1 & 0 & 0 & 0 & 0 & 0 & 0 \\
\hline
 0 & 0 & 0 & 0 & 0 & 0 &  0 &  0 &-1    & 3 &-1 & 0 &-1 & 0 &-1 & 0 \\
 0 & 0 & 0 & 0 & 0 & 0 &  0 &  0 & 0    &-1 & 2 &-1 & 0 & 0 & 0 & 0 \\
 0 & 0 & 0 & 0 & 0 & 0 &  0 &  0 & 0    & 0 &-1 & 2 &-1 & 0 & 0 & 0 \\
 0 & 0 & 0 & 0 & 0 & 0 &  0 &  0 & 0    &-1 & 0 & 0 & 2 &-1 & 0 & 0 \\
 0 & 0 & 0 & 0 & 0 & 0 &  0 &  0 & 0    & 0 & 0 & 0 &-1 & 2 & 0 & 0 \\
 0 & 0 & 0 & 0 & 0 & 0 &  0 &  0 & 0    &-1 & 0 & 0 & 0 & 0 & 2 &-1 \\
 0 & 0 & 0 & 0 & 0 & 0 &  0 &  0 & 0    & 0 & 0 & 0 & 0 & 0 &-1 & 2 \\
\end{array}
\right )
\nonumber
\end{eqnarray}
}
One can convince oneself
that the determinant of this matrix is equal zero and the principal 
minors are positive definite:
The Berger matrix can be considered of built from three diagonal blocks, distinguished by lines,
containing respectively parts which partially resemble
(in fact they only  differ by the apparition of 3's in the joining nodes)  $E_6^{(1)}$ and $A_r$ Cartan matrix copies.

Let construct for this graph
the ¨root¨ system consisting from the
the $7_L+2_{int}+7_R=16$ nodes (see Fig. \ref{fig6}). 
The 16 roots are denoted as 
$\{ \alpha_i,\ldots,{\hat \alpha_1},{\hat \alpha_2},
{\tilde \alpha_i},\ldots\}, i=1-7$. In a similar 
way the $16+3$ dimensional 
orthonormal canonical basis is divided in three sectors and 
given by
$\{ e_i,\ldots, {\hat e_{1}},{\hat e_0},{\hat e_{-1}},{\tilde \alpha_i},
\ldots\}, i=1-8$. 

\begin{eqnarray}
\alpha_1   &=& -\frac{1}{2}(-e_1+e_2+e_3+e_4+e_5+e_6+e_7-e_8)\nonumber\\      
\alpha_2   &=&      (e_2-e_1)                             \nonumber\\
\alpha_3   &=&      (e_4-e_3)                             \nonumber\\
\alpha_4   &=&      (e_5-e_4)                             \nonumber\\
\alpha_5   &=&      (e_1+e_2)                             \nonumber\\
\alpha_6   &=& -\frac{1}{2}(e_1+e_2+e_3+e_4+e_5-e_6-e_7+e_8)  \nonumber  \\   
\alpha_7   &=& -e_2+e_3-{\hat e_1}                          \nonumber  \\
\hat \alpha_1 &=& \hat e_1-\hat e_0                      \nonumber  \\
\hat \alpha_2 &=& \hat e_0-\hat e_{-1}                   \nonumber  \\
\tilde \alpha_7  &=& -\tilde e_2+\tilde e_3-{\hat e}_{-1}  \nonumber  \\  
\tilde \alpha_6  &=& -\frac{1}{2}(\tilde e_1+\tilde e_2+\tilde e_3
+\tilde e_4+\tilde e_5-\tilde e_6-\tilde e_7+\tilde e_8)                 \nonumber  \\  
\tilde \alpha_5   &=&      (\tilde e_1+\tilde e_2)  \nonumber  \\
\tilde \alpha_4   &=&      (\tilde e_5-\tilde e_4)  \nonumber  \\
\tilde \alpha_3   &=&      (\tilde e_4-\tilde e_3)  \nonumber  \\
\tilde \alpha_2   &=&      (\tilde e_2-\tilde e_1)  \nonumber  \\
\tilde \alpha_1   &=& -\frac{1}{2}(-\tilde e_1+\tilde e_2+\tilde e_3
+\tilde e_4+\tilde e_5+\tilde e_6+\tilde e_7-\tilde e_8)                 \nonumber        
\end{eqnarray} 

The matrix has a null eigenvector with entries, the generalized 
Coxeter labels, 
$$\mu=(1,2,1,2,1,2,3,3,3,3,2,1,2,1,2,1 ).$$ 
These numbers are shown attached to the nodes of the generalized 
Dynkin graph in Fig.\ref{fig6}.
The Coxeter number is $h=30$.
Note that the Coxeter labels corresponding to one sector $E_6^{(1)}$ are
precisely a subset of those $=(1,2,1,2,1,2,3)$. 
The following linear combination of the ¨roots¨ satisfies the affine 
condition: 
\begin{eqnarray}
\alpha_{1} + 2 \alpha_{2}+ \alpha_{3}+2\alpha_{4}+\alpha_{5}+ 2\alpha_{6}+ 3\alpha_{7}     
+ 3 \hat\alpha_{1}  +     \hat\alpha_{2}  + 
\tilde
\alpha_{1} + 2 \tilde\alpha_{2}+ \tilde\alpha_{3}+2\tilde\alpha_{4}+\tilde\alpha_{5}+ 2\tilde\alpha_{6}+ 3\tilde\alpha_{7}=0. \nonumber    
 \end{eqnarray}

Note that, as it could not be otherwise, the affine condition entangles all the vectors of the system:
there is only one affine condition, the individual affine conditions 
corresponding to the individual sectors are not satisfied.

Generalizations of 
the $E_7^{1}$
and  $E_8^{1]}$ series of graphs, 
corresponding respectively to vectors (112) and (113) can 
also be easily obtained \cite{newtorrente}.
It is suggestive to consider the 
 procedure outlined in this section as a new way of unifying two 
semisimple algebras within a larger algebra. 
By the reverse process, the decomposition of the Berger diagram leads to 
 proper, Kac-Moody, 
subdiagrams and the reduction of the Berger matrix to Kac-Moody blocks.
This can be performed eliminating the  nodes corresponding
to one or more roots and the lines attached to them.

In particular, it is suggestive to consider the case where a Berger 
matrix resembling by blocks  (procedure exemplified in Fig.\ref{fig6})
$$G\sim CY3B(8_L+2_{int}+8_R)\sim E_7^{(1)} + A_r+ E_7^{(1)}$$ 
or
$$G\sim CY3B(9_L+2_{int}+9_R)\sim E_8^{(1)} + A_r+ E_8^{(1)}$$ 
is reduced to standard blocks and where the
one affine condition or central vector is broken down to 
three different affine conditions for each of the blocks. We could  compare 
this case with the heterotic  string example where the unified gauge group is 
obtained by standard direct sum of algebras $G=E_8+E_8$.

\FIGURE{
\parbox{\linewidth}{
\hspace{-4cm}
 \unitlength0.9mm

\hspace{1cm}

\begin{tabular}{cc}
   \begin{picture}(50,50)
   \multiput(0,0)(10,0){8}{\circle*{1.7}}
   \put(0,0){\line(1,0){70}}
   \multiput(20,-20)(0,10){5}{\circle*{1.7}}
   \put(20,-20){\line(0,1){40}}
   \multiput(50,-20)(0,10){5}{\circle*{1.7}}
   \put(50,-20){\line(0,1){40}}
   \put(20,0){\circle{3}}
   \put(50,0){\circle{3}}
   \put(1,2){1}
   \put(11,2){2}
   \put(21,2){3}
   \put(31,2){3}
   \put(41,2){3}
   \put(51,2){3}
   \put(61,2){2}
   \put(71,2){1}
   \put(21,-18){1}
   \put(21,-8){2}
   \put(21,2){3}
   \put(21,12){2}
   \put(21,22){1}
   \put(51,-18){1}
   \put(51,-8){2}
   \put(51,2){3}
   \put(51,12){2}
   \put(51,22){1}

   \end{picture} & \hspace{2cm}
\quad\quad$\Longrightarrow $
$2\times (\quad $

 \unitlength0.5mm
   \begin{picture}(50,50)
   \multiput(0,0)(10,0){5}{\circle*{1.7}}
   \put(0,0){\line(1,0){40}}
   \multiput(20,0)(0,10){3}{\circle*{1.7}}
   \put(20,0){\line(0,1){20}}
   
    \put(1,2){1}
    \put(11,2){2}
    \put(21,2){3}
    \put(31,2){2}
    \put(41,2){1}
    \put(21,2){3}
    \put(21,12){2}
    \put(21,22){1}

   \end{picture}

$)\quad  +\quad$

\unitlength0.5mm
  \begin{picture}(50,50)
   \multiput(0,0)(10,0){2}{\circle*{1.7}}
   \put(0,0){\line(1,0){10}}
   \multiput(1,2)(10,0){2}{1}
   \end{picture}  

\end{tabular}

}\vspace{2.cm}
\label{fig6}
\caption{
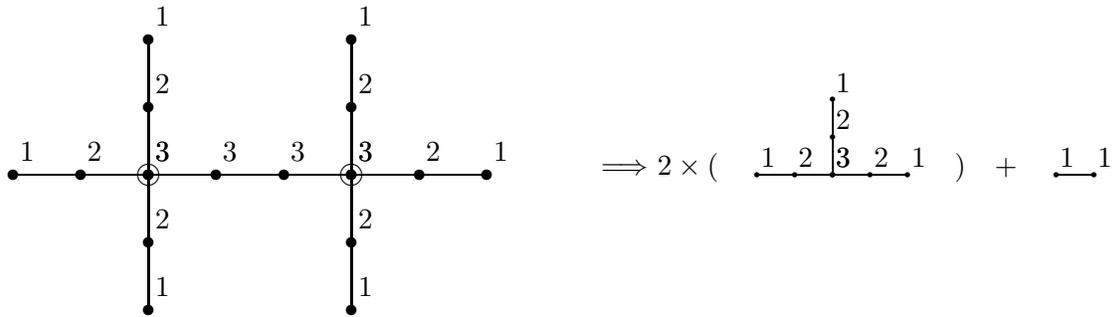
Berger-Dynkin diagrams for  $CY3B(7+2+7)$ diagram and 
breaking down  into two copies of affine  $E_6^{(1)}$ diagrams.}
}

\section{Summary, additional comments and conclusions}

The interest to look for new algebras  beyond Lie algebras started 
from   the $SU(2)$- conformal theories 
(see for example \cite{CIZ,FZ}).
One can think that  geometrical concepts, in particular algebraic geometry,
 could be a  natural and more promising way to  do this.
This marriage of algebra and geometry has been useful in both ways. 
Let us remind that to prove mirror symmetry of Calabi-Yau spaces, 
the greatest  progress was reached with using the technics of Newton reflexive polyhedra  in \cite{Bat}.

 
We considered  here  examples of 
graphs from $K3=CY_2$ 
and new graphs from $CY_d$ ($d \geq 3$) 
reflexive polyhedra, which could be in one-to-one correspondence to 
some kind of generalized  algebras, 
like it was in $K3$ case.
From the large number of possibly graphs 
we illustrated  some selected cases
corresponding 
generalized  symmetric Berger matrices.
It  is very remarkable  that some of these graphs can naturally be
extended into infinite series while some others seem to remain 
``exceptional''.


What would be this new algebraic structure?, which would be the 
set of symmetries associated to this algebra?. 
It is very  well known, by the Serre theorem,
  that Dynkin diagrams defines  one-to-one   Cartan matrices.
In this work, we have generalized    
 some of the properties of Cartan matrices 
for Cartan-Lie and Kac-Moody algebras  into a new class of affine, and non-affine Berger matrices.
We arrive then to the obvious conclusion that 
any algebraic structure emerging from this 
can not be a CLA or KMA  algebra. 

The main difference with previous definitions being in the values that 
diagonal elements of the matrices can take. In Calabi-Yau CY3 spaces, 
new entries with  norm equal to 3 are allowed.
The choice of this number can be related to two facts:  
First, we should take  in mind that in higher 
dimensional Calabi-Yau spaces resolution of singularities should be 
accomplished by more topologically complicated projective spaces:
while for resolution of 
quotient singularities in K3 case one should use the $CP^1$ with Euler 
number 2,
the, Euler number 3,  $CP^2$ space  could 
be use for resolution of some  singularities in $CY_3$ space. 
The second fact is related to the 
cubic matrix theory\cite{Kerner}, where a ternary operation is defined and 
in which the $S_3$ group  naturally appears. 
One conjecture, draft from the fact of the underlying UCYA construction, 
is that, as  
Lie and affine Kac-Moody algebras are based  on 
a binary composition law; 
the emerging picture from the consideration of 
these graphs could lead us to algebras  
including simultaneously different n-ary 
composition rules.
Of course, the underlying UCYA construction could manifest in other 
ways: for example in giving a  framework for a higher level linking 
of algebraic structures: Kac-Moody algebras among themselves and with
any other hypothetical algebra generalizing them.
Thus, putting together  UCYA theory and graphs from reflexive polyhedra,
we expect that 
iterative application of 
non-associative n-ary operations 
give us  not only a complete picture  of the RWV, but allow us in 
addition to establish
``dynamical'' links among
 RWV vector  and graphs  of  different dimensions and, in a further 
step,
links between  singularity blow-up and possibly new
generalized physical
symmetries.

As it is mentioned in the last section, generalizations of 
the $E_7^{1}$
and  $E_8^{1]}$ series of graphs  can 
also be easily obtained.
It is suggestive to consider the 
 procedure outlined in this section as a new way 
of unifying two 
semisimple algebras within a larger algebra. 
By the reverse process, the decomposition of the Berger diagram leads to 
 proper, Kac-Moody, 
subdiagrams and the reduction of the Berger matrix to Kac-Moody blocks.
This can be performed eliminating the  nodes corresponding
to one or more roots and the lines attached to them.

In particular, it is suggestive to consider the case where a Berger 
matrix resembling by blocks copies of $E_7^{(1)},E_8^{(1)}$ matrices
$$G\sim CY3B(8_L+2_{int}+8_R)\leftarrow E_7^{(1)} + A_r+ E_7^{(1)}$$ 
or
$$G\sim CY3B(9_L+2_{int}+9_R)\leftarrow E_8^{(1)} + A_r+ E_8^{(1)}$$ 
are reduced to standard blocks. As a result of 
 this reduction  the single
o affine condition or central vector is broken down to 
three different affine conditions for each of the blocks. We could  compare 
this case with the heterotic  string example where the unified gauge group is 
obtained by standard direct sum of algebras $G=E_8+E_8$.

\vspace{0.6cm}
{\bf Acknowledgments}.
One of us, {G.G.}, would like to give his thanks to E. Alvarez, 
P. Auranche, R. Coquereaux, N. Costa, C. Gomez, B. Gavela, L. Fellin,  A. Liparteliani, 
L.Lipatov,  A. Sabio Vera, J. Sanchez Solano,
P.Sorba,  G. Valente  and  John Ellis
for valuable discussion and nice support.
We  acknowledge the  financial  support of 
 the  Spanish CYCIT  funding agency  and the CERN 
Theoretical Division.

\newpage

{

}
\end{document}